\documentclass[prd,aps,preprint,showpacs,nofootinbib,eqsecnum,superscriptaddress]{revtex4-1}

\usepackage{hyperref}
\usepackage{amsfonts,amssymb,amsmath,amsthm,amsxtra,mathtools,euscript,mathrsfs}

\usepackage{bm}
\usepackage{graphicx,color,xcolor,subfigure}
\usepackage{dcolumn}
\usepackage{bbm}

\def\a{\alpha}

\def\d{\delta}

\def\ti{\tilde}

\begin{document}
\title{Modified Cosmology Models from Thermodynamical Approach}

\author{Chao-Qiang Geng}
\email[Electronic address: ]{geng@phys.nthu.edu.tw}
\affiliation{School of Fundamental Physics and Mathematical Sciences\\Hangzhou Institute for Advanced Study, UCAS, Hangzhou, 310024 China }
\affiliation{Department of Physics,
National Tsing Hua University, Hsinchu 300}
\affiliation{Physics Division,
National Center for Theoretical Sciences, Hsinchu 300}
\author{Yan-Ting Hsu}
\email[Electronic address: ]{ythsu@gapp.nthu.edu.tw}
\affiliation{Department of Physics,
National Tsing Hua University, Hsinchu 300}
\author{Jhih-Rong Lu}
\email[Electronic address: ]{jhih-ronglu@gapp.nthu.edu.tw}
\affiliation{Department of Physics,
National Tsing Hua University, Hsinchu 300}
\author{Lu Yin}
\email[Electronic address: ]{yinlu@gapp.nthu.edu.tw}
\affiliation{Department of Physics,
National Tsing Hua University, Hsinchu 300}
\begin{abstract}
We apply the first law of thermodynamics to the apparent horizon of the universe with the power-law corrected and
non-extensive Tsallis entropies rather than the Bekenstein-Hawking one. We examine the cosmological properties in the two entropy models by using the CosmoMC package. In particular,  the first numerical study for the cosmological observables with the power-law corrected entropy
is performed. We also show that the neutrino mass sum has a non-zero central value with a relaxed upper bound in the  Tsallis entropy model
comparing with that in the $\Lambda$CDM one.
\end{abstract}

\maketitle

\section{Introduction}
According to the current cosmological observations, our universe
is experiencing a late time accelerating expansion.
Although the $\Lambda$CDM model can describe the accelerating
universe by introducing dark energy~\cite{Amendola:2015}, it fails to
solve the cosmological constant problem,
related to the ``fine-tuning"~\cite{Weinberg:1988cp, Weinberg:1972} and ``coincidence"~\cite{ArkaniHamed:2000tc, Peebles:2002gy} puzzles. A lot of efforts have been made to understand these issues.
For example, one can modify the gravitational theory to
obtain viable cosmological models with dynamical dark energy
to explain the accelerating universe~\cite{Copeland:2006wr}.

On the other hand, one can reconstruct the Friedmann equations through 
the implications of thermodynamics. It has been shown that
the Einstein's equations can be derived by considering
the Clausius' relation of a local Rindler observer
~\cite{Jacobson:1995ab}. In particular, this idea
has been applied to cosmology,
while the Friedmann equations have been obtained by using the 
first law of thermodynamics in the horizon of the universe~\cite{Cai:2005ra}.
It has been also demonstrated that the modified Friedmann equations  
can be acquired from the thermodynamical approach
by just replacing the entropy-area relation with a proper one in a wide variety of gravitational theories~\cite{Cai:2005ra,Akbar:2006er,Akbar:2006kj,Jamil:2009eb,Fan:2014ala,Gim:2014nba}.
Thus, as long as there is a new entropy area relation, 
thermodynamics gives us a new way to determine the
modified Friedmann equations without knowing the 
underlying gravitational theory.
 Furthermore, since the entropy area relation obtained from the modified gravity theory can be useful to extract the dark energy dynamics along with the modified Friedmann  equations, it is reasonable to believe that even if we do not know the underlying theory of modified gravity, some modifications of 
 the entropy relation will still give us additional  information for modified Friedmann equations as well as the dynamics of dark energy, which would be different from $\Lambda$CDM.
As a result,  we expect that  the modification of the entropy is also relevant to the cosmological evolutions.

It is known  that a power-law corrected term from the quantum entanglement can be included in the black hole entropy near its horizon~\cite{Das:2007mj}.
Interestingly, one can apply it to cosmology by taking it
as the entropy on the horizon of the universe.
On the other hand,
the universe is regarded as a non-extensive thermodynamical
system, so the Boltmann-Gibbs entropy should be
generalized to a non-extensive quantity, the Tsallis entropy, while the standard one can be treated as a limit~\cite{Tsallis:1987eu, Lyra:1998wz, Wilk:1999dr}.
The Tsallis entropy has been widely discussed in the literature.
In the entropic-cosmology scenario~\cite{Easson:2010av},
the Tsallis entropy model predicts a decelerating and accelerating
universe~\cite{Komatsu:2013qia}. In addition, a number of works on the Tsallis holographic dark energy have been proposed and investigated~\cite{Abreu:2014ara}. In addition, the Tsallis entropy has also been used in many different dark energy models, such as 
the Barboza-Alcaniz and Chevalier-Polarski-Linder parametric dark energy and  Wang-Meng and Dalal vacuum decay models~\cite{Nunes:2014jra}.
Moreover, it is  shown that modified cosmology from the first law of thermodynamics with varying-exponent Tsallis entropy can  provide a description of both inflation and late-time acceleration with the same parameter choices~\cite{Nojiri:2019skr}.
In particular,
the Tsallis entropy is proportional to a power of the horizon 
area, i.e. $S_T \propto A^{\delta}$, when the universe is assumed to be a spherically symmetric system
~\cite{Tsallis:2012js}.

Although  it is possible to modify Friedmann  equations by just considering fluid with an inhomogeneous equation of state of the corresponding form~\cite{1205.3421}, we still choose the thermo-dynaimical approach
as that in Ref.~\cite{Lymperis:2018iuz}, in which
 the authors considered the first law of thermodynamics of the universe with fixed-exponent Tsallis entropy
and showed that the cosmological evolution mimics that of $\Lambda$CDM
and are in great agreement with Supernovae type Ia observational
data.
In this paper, we examine the features of the modified Friedmann
equations obtained by replacing the usual
Bekenstein-Hawking entropy-area relation,
$S=A/4G$, with the power-law correction
and Tsallis entropies~\cite{Das:2007mj, Tsallis:1987eu, Lyra:1998wz, Wilk:1999dr,
Komatsu:2013qia, Abreu:2014ara, Nunes:2014jra, Zadeh:2018wub, Nojiri:2019skr, Tsallis:2012js, Lymperis:2018iuz}, where $G$ is the gravitational constant.

This paper is organized as follows. In Sec.~\uppercase\expandafter{\romannumeral 2}, we consider the
power-law corrected and Tsallis entropy models and derive
the modified Friedmann equations and dynamical equation of state parameters by applying the first law of thermodynamics to the
apparent horizon of the universe.
In Sec.~\uppercase\expandafter{\romannumeral 3},
we present the cosmological evolutions of the two models and compare them with those in $\Lambda$CDM.
Finally, the conclusions are given in Sec.~\uppercase\expandafter{\romannumeral 4}.
The paper is written in  units of $c=\hbar=k_B=1$.

\section{The Models}
We use the flat Friedmann-Lema\^{i}tre-Robertson-Walker
(FLRW) metric:
\begin{equation}
ds^2=-dt^2+a^{2}(t)\Big(dr^2+r^2 d\Omega^2\Big),
\end{equation}
where $a(t)$ is the scale factor.
The modified Friedmann equations can be constructed by considering
the first law of thermodynamics in the apparent horizon
of the universe and using the new entropy area relation
rather than the Bekenstein-Hawking one.
We concentrate on two models:
power law corrected entropy (PLCE)
and Tsallis entropy cosmological evolution (TECE) Models.
\subsection{Power Law Corrected Entropy (PLCE) Model}
In the PLCE model, the entropy
has the form~\cite{Das:2007mj}
\begin{align}
\label{eq:splce}
S_{pl}= \frac{A}{4L_p^2}\bigg(1 - K_\nu A^{1-\frac{\nu}{2}}\bigg),
\end{align}
where $\nu$ is a dimensionless constant parameter and $K_\nu=\nu (4 \pi)^{(\nu-2)/2}(4-\nu)^{-1}r_c^{\nu-2}$ with $r_c$ the crossover scale,
$A$ corresponds to the area of the system, and $L_p$ represents the Planck length. With the method described in Ref.~\cite{Lymperis:2018iuz},
one is able to extract the modified Friedmann equations:
\begin{align}
H^2&=\frac{8\pi G}{3}(\rho_m +\rho_r + \rho_{DE}),\nonumber\\
\dot{H}&=-4\pi G(\rho_m +\rho_r+ \rho_{DE}+p_m +p_r+p_{DE}),\label{FeqTsa}
\end{align}
where $\rho_{DE}$ and $p_{DE}$ are the dark energy density
and pressure, given by
\begin{align}
\rho_{DE} &= \frac{3}{8 \pi G}\frac{1}{r^{2-\nu}_c}\big(H^{\nu}-1\big)
+ \frac{\Lambda}{8 \pi G},\\
p_{DE} &=\frac{-\nu}{8 \pi G}\frac{\dot{H}}{r^{2-\nu}_c}\big(H^{\nu}-1\big)- \frac{3}{8 \pi G}\frac{1}{r^{2-\nu}_c}\big(H^{\nu}-1\big)
-\frac{\Lambda}{8 \pi G},
\end{align}
respectively.
To discuss the evolution of dark energy, it is convenient to
define the equation of state parameter, $w_{DE} \equiv p_{DE}/\rho_{DE}$,
which is found to be
\begin{align}
w_{DE} = -1 + \frac{-\nu\dot{H}H^{\nu-2}}{3(H^{\nu}-1)
	+ \Lambda r^{2-\nu}_c}.
\end{align}
\subsection{Tsallis Entropy Cosmological Evolution Model}
In the TECE model, we  have~\cite{Tsallis:2012js}
\begin{equation}
\label{eq:stece}
S_T=\frac{\ti{\alpha}}{4G}A^{\delta},
\end{equation}
where $A$ is the area of the
system with dimension [$L^2$], $\ti{\alpha}$ is a
positive constant with dimension [$L^{2-2\delta}$], and 
$\delta$ denotes the non-additivity parameter.
Similarly, by following the procedure in Ref.~\cite{Lymperis:2018iuz},
we obtain
\begin{align}
H^2&=\frac{8\pi G}{3}(\rho_m +\rho_r + \rho_{DE}),\nonumber\\
\dot{H}&=-4\pi G(\rho_m +\rho_r+ \rho_{DE}+p_m +p_r+p_{DE}),\label{FeqTsa}
\end{align}
with
\begin{align}
\rho_{DE}&=\frac{3}{8 \pi G}\bigg[\frac{\Lambda}{3}+
H^2\bigg(1-\alpha \frac{\delta}{2-\delta} H^{2(1-\delta)}\bigg)\bigg],\\
p_{DE}&=-\frac{1}{8 \pi G}\bigg[\Lambda 
+2 \dot{H}(1-\a\d H^{2(1-\d)})
+3H^2\bigg(1-\alpha\frac{\d}{2-\d}H^{2(1-\d)}\bigg)\bigg]
\end{align}
where $\alpha= (4 \pi)^{\delta-1}\ti{\alpha}$, and $\Lambda$ is a constant related to the present values of $H_0, \rho_{m0}$ and $\rho_{r0}$, given by
\begin{align}
\Lambda&=\frac{3\alpha\delta}{2-\delta}H_0^{2(2-\delta)}-8\pi G(\rho_{m0}
+\rho_{r0}).
\end{align}
Thus, the equation of state parameter for the TECE model is evaluated to be
\begin{align}
w_{DE} = \frac{p_{DE}}{\rho_{DE}}
=-1 + \frac{2\dot{H}(\a \d H^{2\dot{H}(1-\d)}-1)}
{3H^2\bigg(1-\frac{\a \d}{2-\d}H^{2(1-\d)}\bigg)+\Lambda}.
\end{align}

\section{Cosmological Evolutions}
\label{sec:observation}

\subsection{Power Law Corrected Entropy Model}
Since $\rho_{DE}$ and $w_{DE}$  are determined by the Hubble parameter $H(z)$, we use the Newton-Raphson method~\cite{press:2007} to obtain the cosmological evolutions of the PLCE model. 

\begin{figure}[h]
	\centering
	\includegraphics[width=0.45 \linewidth, angle=270]{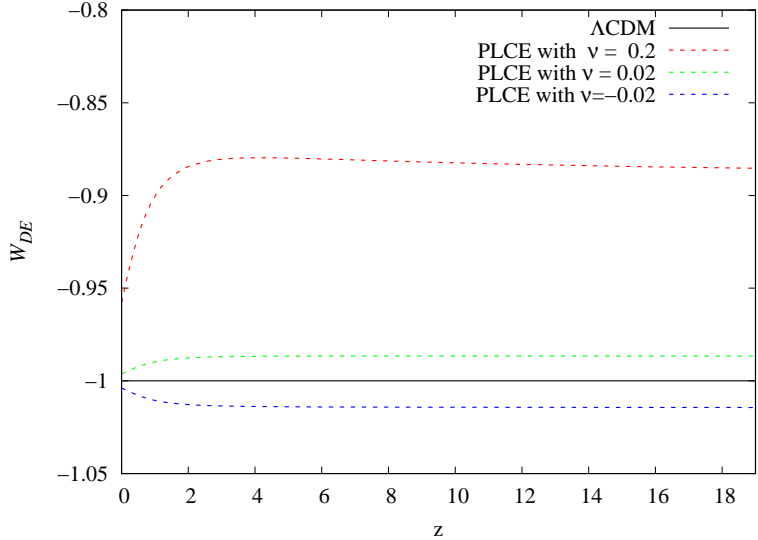}
	\caption{Evolutions of the equation-of-state parameter $w_{DE}$ in $\Lambda$CDM and PLCE models.} 
	\label{fg:W1}
\end{figure}

Because the PLCE model goes back to $\Lambda$CDM when $\nu = 0$, we choose $\nu =\pm0.02$ to compare the differences between the two models. We also take a larger value of $\nu = 0.2$ to check the sensitivity of $\nu$. The results in Fig.~\ref{fg:W1} show that $w_{DE}$ does not overlap or cross -1 in any non-zero value of $\nu$. In addition, it maintains its value in the early universe, and only trends to -1 for $z<2$.

\begin{figure}[b]
	\centering
	\includegraphics[width=0.45 \linewidth, angle=270]{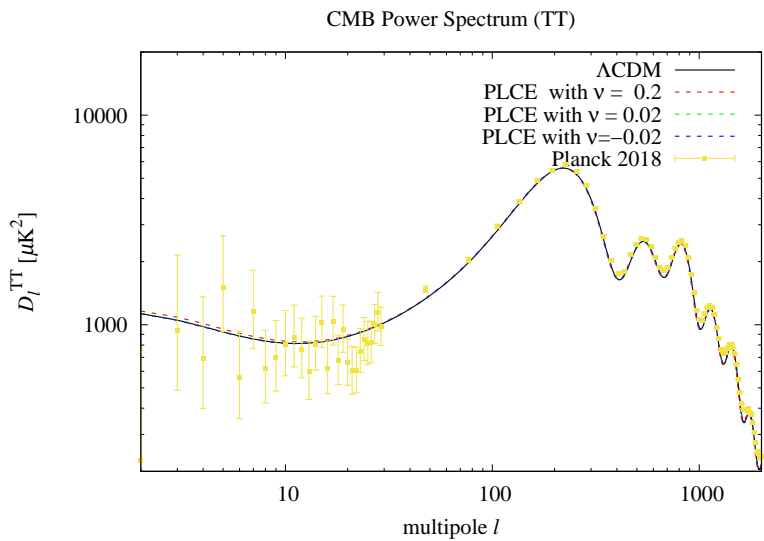}
	\caption{CMB power spectra of the TT mode in $\Lambda$CDM and PLCE models along with the Planck 2018 data.}
	\label{fg:TT1}
\end{figure}

\begin{figure}[h]
	\centering
	\includegraphics[width=0.45 \linewidth, angle=270]{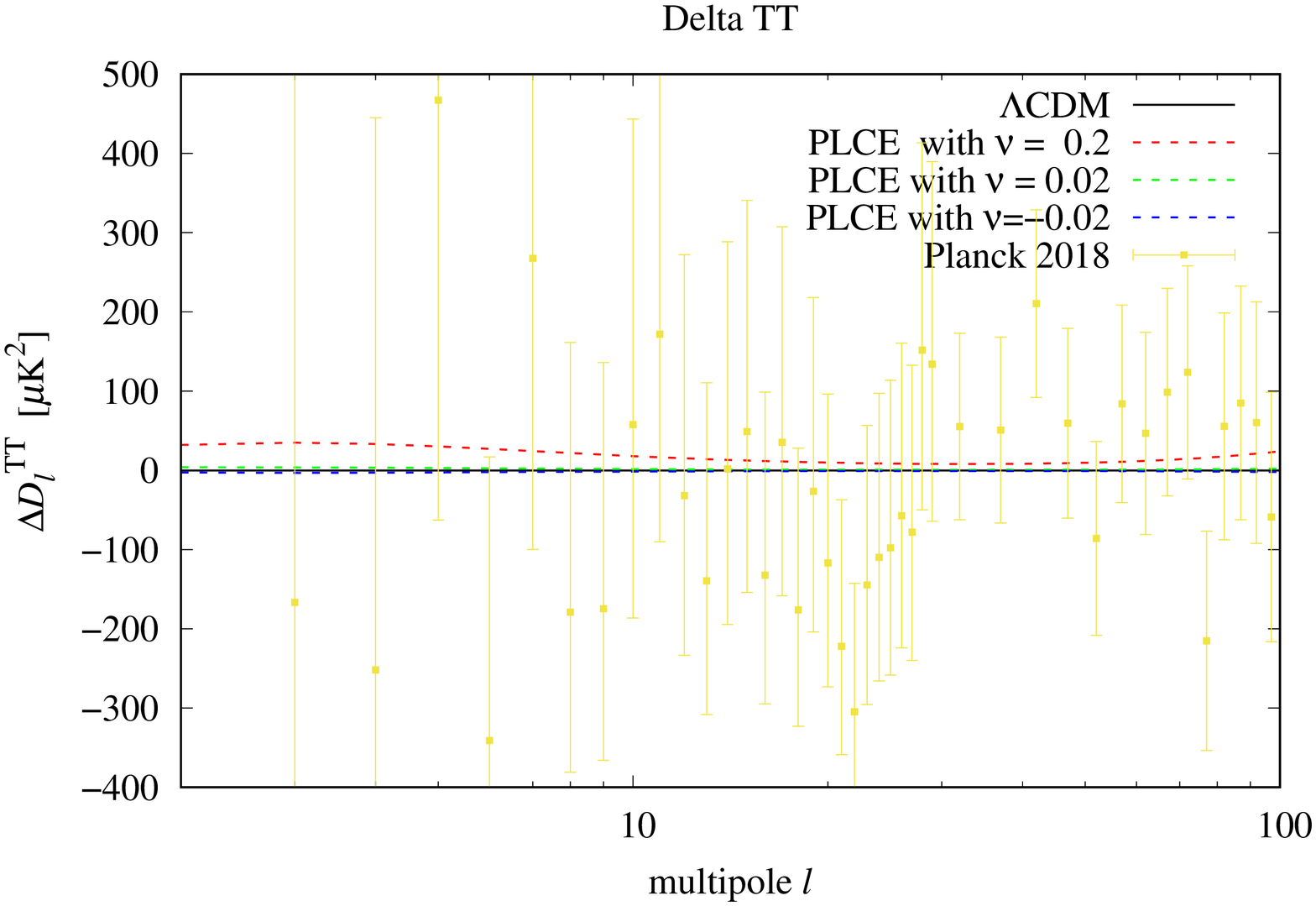}	
	\caption{
		The change $\Delta D_{\ell}^{TT}$ of the TT mode of CMB power spectra between PLCE and $\Lambda$CDM, where the legend is the same as Fig.~\ref{fg:W1}.}
	\label{fg:DIFTT1}
\end{figure}

In Fig.~\ref{fg:TT1}, we display the CMB power spectra in the $\Lambda$CDM and PLCE models
along with the data from Planck 2018.
Since the TT  spectra of PLCE and $\Lambda$CDM are almost identical to the data from Planck 2018 for the high values of the multipole $l$, 
we focus on the differences between the two models and the data when $l<100$ as depicted in Fig.~\ref{fg:DIFTT1}. The TT power spectrum in the PLCE model for $\nu > 0$ is larger than that of $\Lambda$CDM  when $l < 100$ with the error in the allowable range of the observational data.

\begin{figure}[h]
	\centering
	\includegraphics[width=0.45 \linewidth, angle=270]{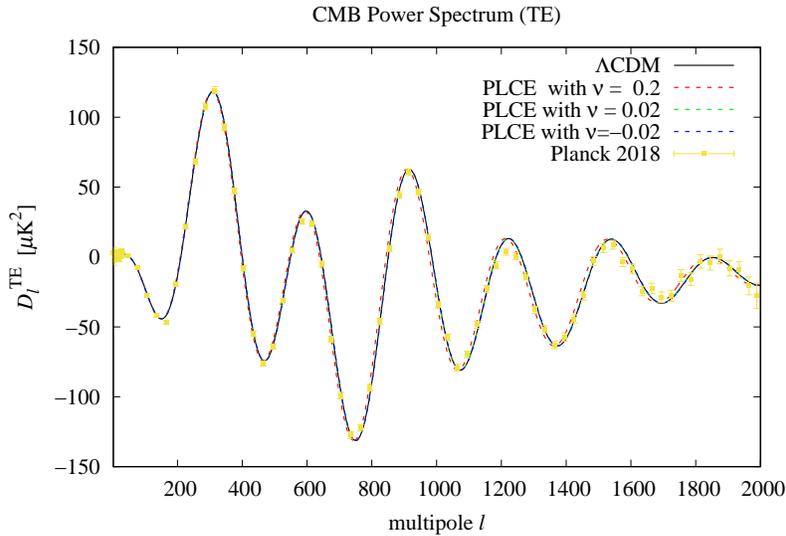}
	\caption{CMB TE power spectra in the $\Lambda$CDM and PLCE models along with the Planck 2018 data.} 
	\label{fg:TE1}
\end{figure}

\begin{figure}[h]
	\centering
	\includegraphics[width=0.45 \linewidth, angle=270]{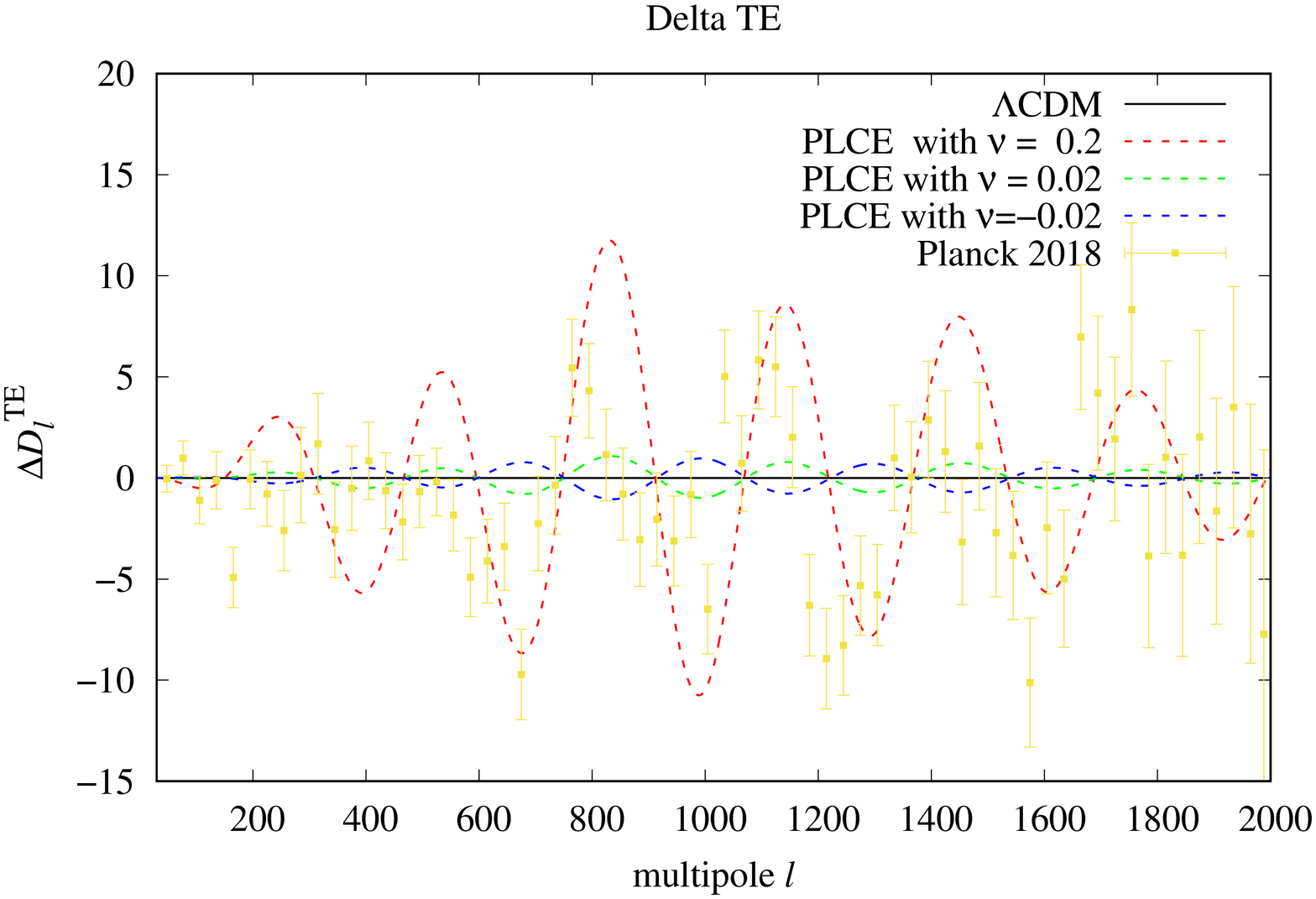}	
	\caption{
		 The change $\Delta D_{\ell}^{TE}$ of the TT mode of CMB power spectra between PLCE and $\Lambda$CDM, where the legend is the
	     same as Fig.~\ref{fg:TE1}.} 
	\label{fg:DIFTE1}
\end{figure}

For the TE mode, the spectra in PLCE for the different parameters $\nu$ are always close to that in $\Lambda$CDM as well as the observational data of Planck 2018, as shown in Fig.~\ref{fg:TE1}. However, when we carefully compare the differences between the results in PLCE and  $\Lambda$CDM in Fig.~\ref{fg:DIFTE1}, we notice that those of PLCE are closer to the Planck 2018 data, comparing to  that in $\Lambda$CDM.

\subsection{Tsallis Entropy Cosmological Evolution Model}


\begin{figure}[h]
	\centering
	\includegraphics[width=0.45 \linewidth, angle=270]{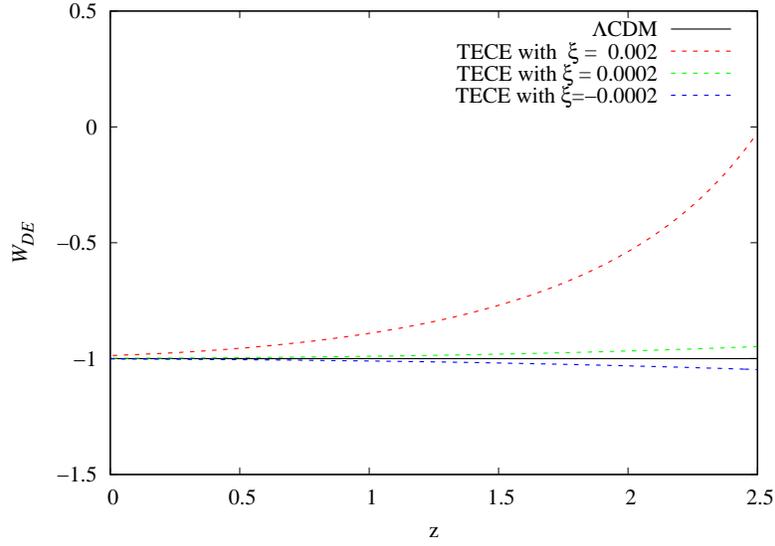}
	\caption{Evolutions of the equation-of-state parameter $w_{DE}$ in $\Lambda$CDM and TECE models.}
	\label{fg:W2}
\end{figure}
	Eq.~\eqref{eq:stece} of TECE becomes the one in $\Lambda$CDM when $\delta=\ti{\alpha}=1$. In our study, we only focus on the effects
	when $\delta \neq 1$ so we set $\ti{\alpha}=1$ and $\delta=1+\xi$. In Fig.~\ref{fg:W2}, we find that the equation of state, $w_{DE}$,  behaves differently for different values of $\xi$. In particular, it is larger (smaller) than -1 when $\xi$ is larger (smaller) than zero without  crossing  -1 in anytime.

\begin{figure}[h]
	\centering
	\includegraphics[width=0.45 \linewidth, angle=270]{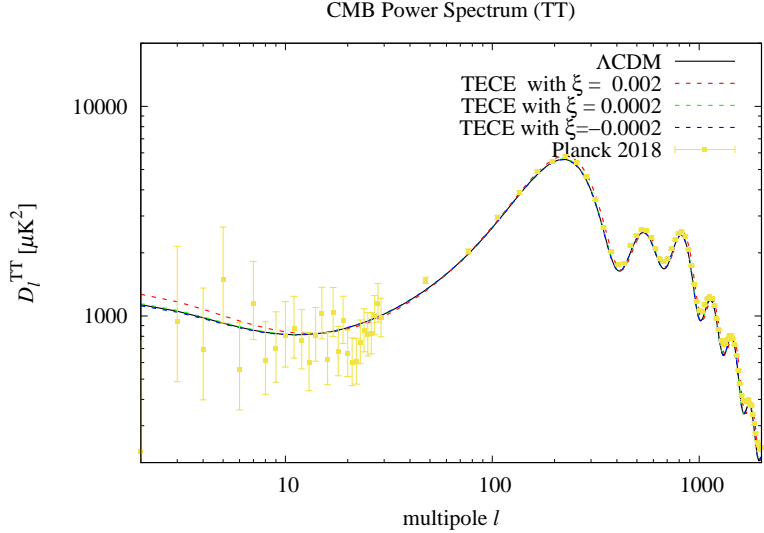}
	\caption{Legend is the same as Fig.~\ref{fg:TT1} but in the TECE model
		with a set of $\xi$.} 
	\label{fg:TT2}
\end{figure}

\begin{figure}[h]
	\centering
	\includegraphics[width=0.45 \linewidth, angle=270]{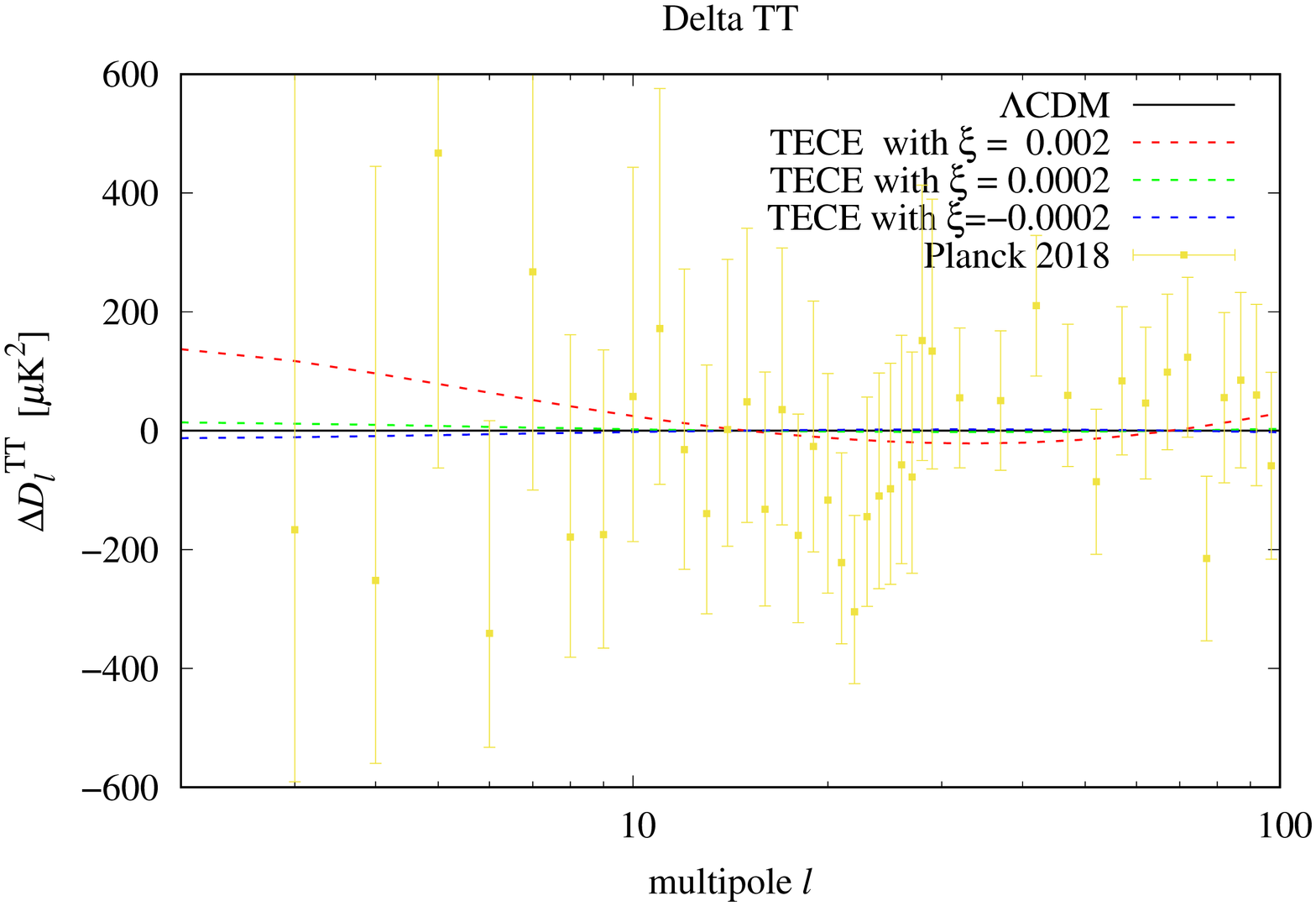}	
	\caption{
		Legend is the same as Fig.~\ref{fg:DIFTT1} but in the TECE model
		with a set of $\xi$.} 
	\label{fg:DIFTT2}
\end{figure}

\begin{figure}[h]
	\centering
	\includegraphics[width=0.45 \linewidth, angle=270]{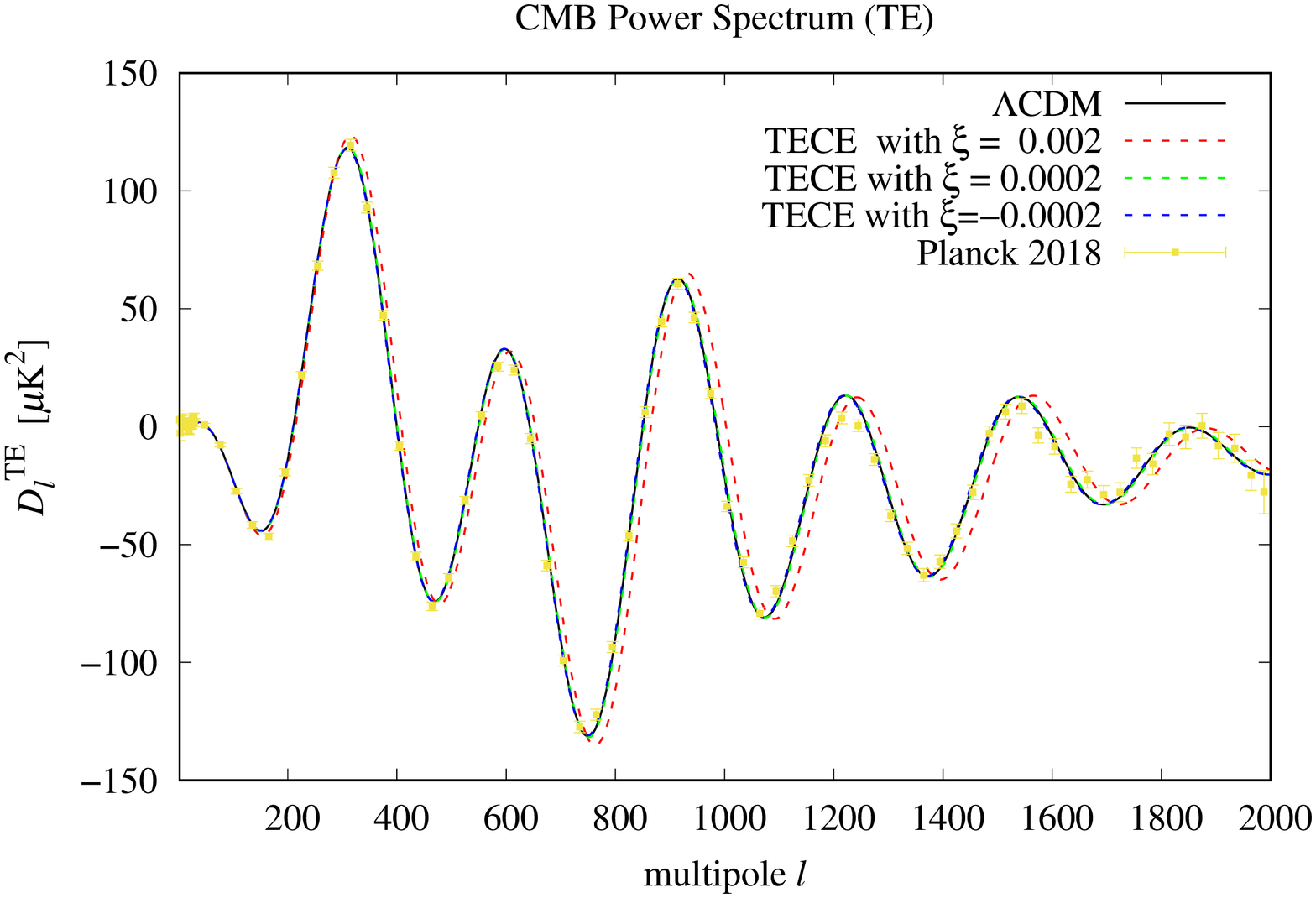}
	\caption{Legend is the same as Fig.~\ref{fg:TE1} but in the TECE model
		with a set of $\xi$.} 
	\label{fg:TE2}
\end{figure}

\begin{figure}[h]
	\centering
	\includegraphics[width=0.45 \linewidth, angle=270]{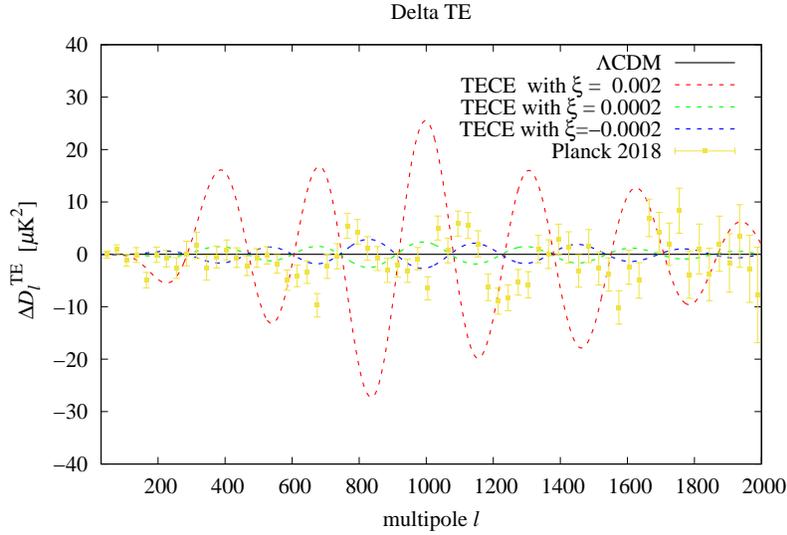}	
	\caption{
		Legend is the same as Fig.~\ref{fg:DIFTE1} but in the TECE model
		with a set of $\xi$.}
	\label{fg:DIFTE2}
\end{figure}

In Figs.~\ref{fg:TT2} and ~\ref{fg:DIFTT2}, we see that the TT Power spectra of TECE and $\Lambda$CDM have a large difference in the large scale structure. Note that there is a significant discrepancy between $\Lambda$CDM and the data at $l\sim 20-27$. However, the spectrum of TECE 
for  $\xi$=0.002 and $l\sim 20-27$ is below that in $\Lambda$CDM, and closer to the observational data of Planck 2018. The shifts of the
TE mode between PLCE and $\Lambda$CDM are shown in Figs.~\ref{fg:TE2} and
\ref{fg:DIFTE2}.

\subsection{Global Fits}

We use the modified ${\bf CAMB}$ and {\bf CosmoMC} program~\cite{Lewis:2002ah} to do the global cosmological
fits for the PLCE and TECE models from the observational data with the MCMC method.
The dataset includes those of the CMB temperature fluctuation from {\it Planck 2015} with TT, TE, EE, low-$l$
polarization and CMB lensing from SMICA~\cite{Adam:2015wua, Aghanim:2015xee, Ade:2015zua}, the weak lensing (WL) data from CFHTLenS ~\cite{Heymans:2013fya}, and the BAO data from 6dF Galaxy Survey~\cite{Beutler:2011hx} and BOSS~\cite{Anderson:2013zyy}.
In particular, we include 35 points for  the $H(z)$ measurements in our fits, which are listed in Table~\ref{tab:0}.
The $\chi^2$ fit is given by
\begin{eqnarray}
\label{eq:chi}
{\chi^2}={\chi^2_{CMB}}+{\chi^2_{WL}}+{\chi^2_{BAO}}+{\chi^2_{H(z)}},
\end{eqnarray}
with
\begin{eqnarray}
\chi^2_c = \sum_{i=1}^n \frac{(T_c(z_i) - O_c(z_i))^2}{E^i_c} \,,
\end{eqnarray}
where the subscript of ``$c$" denotes the category of the data, $n$ represents the number of the  dataset, 
$T_c$ is the prediction from {\bf CAMB}, and $O_c$ ($E_c$) corresponds to the observational value (covariance). The priors of the various 
cosmological parameters are given in Table~\ref{tab:1}.

\begin{table}[!hbp]
	\caption{$H(z)$ data points}
	\begin{tabular}{|c|c|c|c||c|c|c|c||c|c|c|c|}
		\hline
		\ & $z$ & $H(z)$ & Ref. & \ & $z$ & $H(z)$ & Ref. & \ & $z$ & $H(z)$ & Ref. \\
		\hline
		~1~ & 0.07 & 69.0$\pm$19.6 & ~\cite{Zhang:2012mp} &
		~13~ & 0.4 & 95.0$\pm$17.0 &~\cite{Simon:2004tf} &
		~25~ & 0.9 & 117.0$\pm$23.0 & ~\cite{Simon:2004tf} \\
		\hline
		2 & 0.09 & 69.0$\pm$12.0 & ~\cite{Jimenez:2003iv} & 14 & 0.4004 & 77.0$\pm$10.2 &~\cite{Moresco:2016mzx} &26 & 1.037 & 154.0$\pm$20.0 & ~\cite{Moresco:2012jh}\\
		\hline
		3 & 0.12 & 68.6$\pm$26.2 & ~\cite{Zhang:2012mp} &15 & 0.4247 & 87.1$\pm$11.2 & ~\cite{Moresco:2016mzx}&27 & 1.3 & 168.0$\pm$17.0 & ~\cite{Simon:2004tf}\\
		\hline
		4 & 0.17 & 83.0$\pm$8.0 & ~\cite{Simon:2004tf} &16 & 0.4497 & 92.8$\pm$12.9 & ~\cite{Moresco:2016mzx}& 28 & 1.363 & 160.0$\pm$33.6 & ~\cite{Moresco:2015cya}\\
		\hline
		5 & 0.179 & 75.0$\pm$4.0 & ~\cite{Moresco:2012jh} &17 & 0.4783 & 80.9$\pm$9.0 & ~\cite{Moresco:2016mzx}& 29 & 1.43 & 177.0$\pm$18.0 &  ~\cite{Simon:2004tf}\\
		\hline
		6 & 0.199 & 75.0$\pm$5.0 & ~\cite{Moresco:2012jh} &18 & 0.48 & 97.0$\pm$62.0 &~\cite{Stern:2009ep} & 30 & 1.53 & 140.0$\pm$14.0 &  ~\cite{Simon:2004tf}\\
		\hline
		7 & 0.2 & 72.9$\pm$29.6 & ~\cite{Zhang:2012mp} &19 & 0.57 & 92.4$\pm$4.5 & ~\cite{Reid:2012sw}&31 & 1.75 & 202.0$\pm$40.0 &  ~\cite{Simon:2004tf} \\
		\hline
		8 & 0.27 & 77.0$\pm$14.0 & ~\cite{Simon:2004tf} &20 & 0.5929 & 104.0$\pm$13.0 & ~\cite{Moresco:2012jh} & 32 & 1.965 & 186.5$\pm$50.4 & ~\cite{Moresco:2015cya}\\
		\hline
		9 & 0.24 & 79.69$\pm$2.65 &  ~\cite{Gaztanaga:2008xz} &21 & 0.6797 & 92.0$\pm$8.0 & ~\cite{Moresco:2012jh}& 33 & 2.3 & 224$\pm$8 & ~\cite{Busca:2012bu}\\
		\hline
		10 & 0.28 & 88.8$\pm$36.6 & ~\cite{Zhang:2012mp} &22 & 0.7812 & 105.0$\pm$12.0 & ~\cite{Moresco:2012jh}&34 & 2.34 & 222$\pm$7 & ~\cite{Hu:2014vua}\\
		\hline
		11 & 0.352 & 83.0$\pm$14.0 & ~\cite{Moresco:2012jh} & 23 & 0.8754 & 125.0$\pm$17.0 & ~\cite{Moresco:2012jh}& 35 & 2.36 & 226$\pm$8 &~\cite{Font-Ribera:2013wce} \\
		\hline
		~12~ & 0.3802 & 83.0$\pm$13.5 & ~\cite{Moresco:2016mzx} &24 & 0.88 & 90.0$\pm$40.0 &~\cite{Stern:2009ep}& & & & \\
		\hline
	\end{tabular}
	\label{tab:0}
\end{table}

\begin{table}[ht]
	\begin{center}
		\caption{ Priors for cosmological parameters in the PLCE and TECE models.  }
		\begin{tabular}{|c||c|} \hline
			Parameter & Prior
			\\ \hline
			PLCE Model parameter $\nu$& $-0.025 \leq \nu \leq  1.0$
			\\ \hline
			TECE Model parameter $\xi$& $-0.01 \leq \xi \leq 0.02$
			\\ \hline
			Baryon density & $0.5 \leq 100\Omega_bh^2 \leq 10$
			\\ \hline
			CDM density & $0.1 \leq 100\Omega_ch^2 \leq 99$
			\\ \hline
			Optical depth & $0.01 \leq \tau \leq 0.8$
			\\ \hline
			Neutrino mass sum& $0 \leq \Sigma m_{\nu} \leq 2$~eV
			\\ \hline
			$\frac{\mathrm{Sound \ horizon}}{\mathrm{Angular \ diameter \ distance}}$  & $0.5 \leq 100 \theta_{MC} \leq 10$
			\\ \hline
			Scalar power spectrum amplitude & $2 \leq \ln \left( 10^{10} A_s \right) \leq 4$
			\\ \hline
			Spectral index & $0.8 \leq n_s \leq 1.2$
			\\ \hline
		\end{tabular}
		\label{tab:1}
	\end{center}
\end{table}

\begin{figure}[h]
	\centering
	\includegraphics[width=0.96 \linewidth]{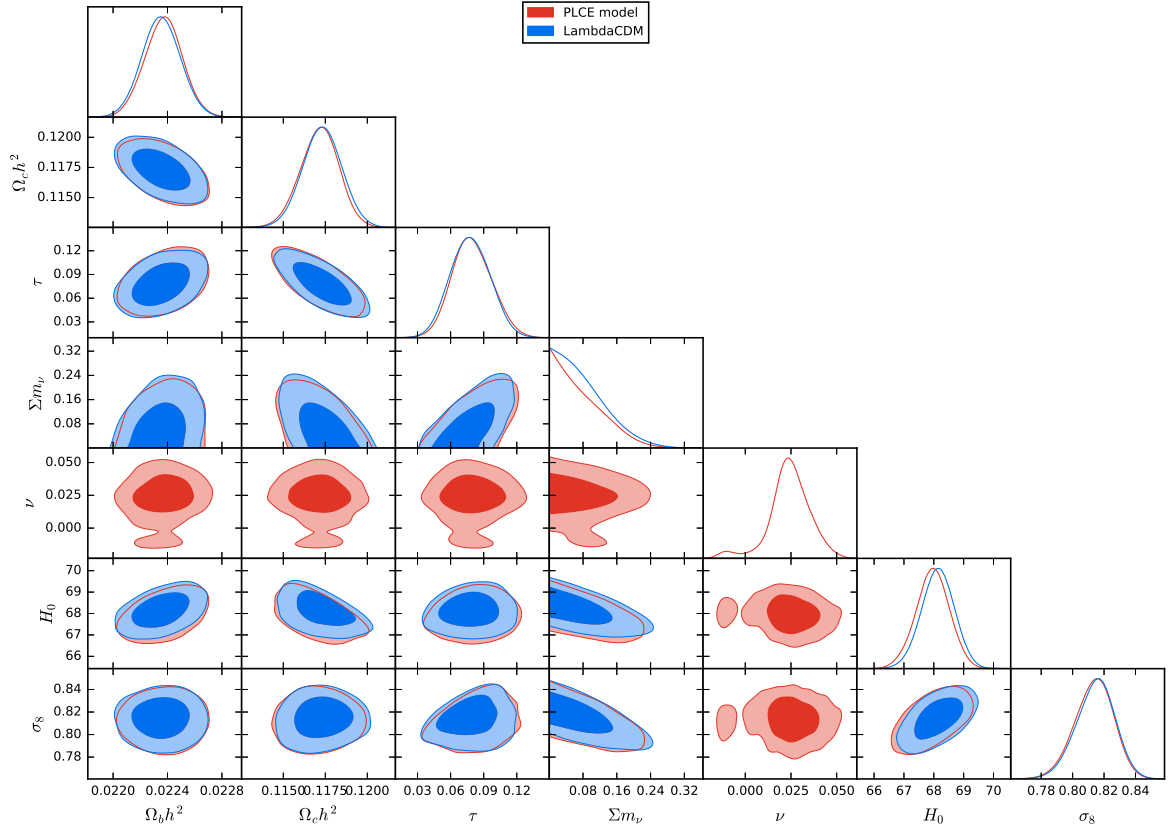}
	\caption{One and two-dimensional distributions of $\Omega_b h^2$, $\Omega_c h^2$, $\sum m_\nu$, $\nu$, $H_0$, and $\sigma_8$ in the PLCE and $\Lambda$CDM  models, where the contour lines represent 68$\%$~ and 95$\%$~ C.L., respectively.}
	\label{fg:P}
\end{figure}

\begin{figure}[h]
	\centering
	\includegraphics[width=0.96 \linewidth]{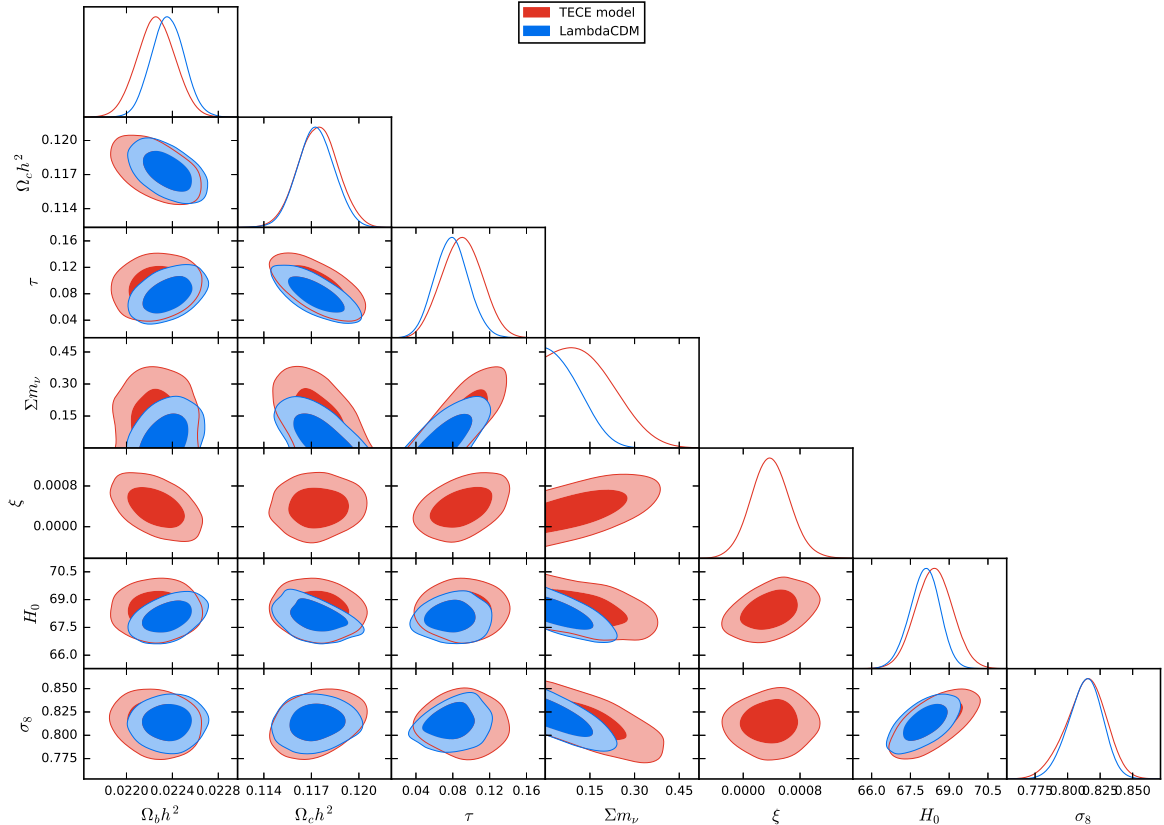}
	\caption{Legend is the same as Fig.~\ref{fg:P} but for the TECE
		and $\Lambda$CDM models.}
	\label{fg:T}
\end{figure}

\begin{table}[h]
	\begin{center}
		\caption{Fitting results for the PLCE and $\Lambda$CDM models, where the limits are given at 68$\%$ and 95$\%$ C.L., respectively }
		\begin{tabular} {|c|c|c|c|c|}
			\hline
			Parameter & PLCE (68\% C.L.)& PLCE (95\% C.L.) & $\Lambda$CDM (68\% C.L.)& $\Lambda$CDM (95\% C.L.)\\
			\hline
            {\boldmath$\Omega_b h^2   $} & $0.02237\pm 0.00014        $& $0.02237 \pm 0.00027 $& $0.02235\pm 0.00014        $& $0.02235^{+0.00028}_{-0.00027}$\\
            
            {\boldmath$\Omega_c h^2   $} & $0.1172^{+0.0012}_{-0.0011}$& $0.1172^{+0.0022}_{-0.0023}$& $0.1173\pm 0.0012          $& $0.1173 \pm 0.0023 $\\
            
            {\boldmath$100\theta_{MC} $} & $1.04101\pm 0.00030        $& $1.04101 \pm 0.00059$& $1.04100\pm 0.00029        $& $1.04100^{+0.00057}_{-0.00058}$\\
            
            {\boldmath$\tau           $} & $0.079^{+0.017}_{-0.019}   $& $0.079^{+0.036}_{-0.034}   $& $0.078\pm 0.018            $& $0.078^{+0.035}_{-0.034}   $\\
            
            {\boldmath$\Sigma m_\nu   $/eV} & $< 0.0982                  $& $< 0.183                   $& $< 0.100                   $ & $< 0.195                   $\\
            
            {\boldmath$\nu            $} & $0.0240^{+0.0110}_{-0.0085} $& $0.024^{+0.022}_{-0.033}   $&$-$&$-$\\
            
            {\boldmath${\rm{ln}}(10^{10} A_s)$} & $3.086^{+0.031}_{-0.035}   $& $3.086^{+0.068}_{-0.063}   $& $3.083\pm 0.033            $& $3.083^{+0.066}_{-0.064}   $\\
            
            $H_0                       $ & $67.96\pm 0.56             $& $68.0 \pm 1.1        $& $68.14\pm 0.55             $& $68.1 \pm 1.1        $\\
            
            $\sigma_8                  $ & $0.814^{+0.013}_{-0.011}   $& $0.814^{+0.023}_{-0.026}   $ & $0.815^{+0.013}_{-0.011}   $& $0.815^{+0.023}_{-0.025}   $\\
            \hline
			$\chi^2_{best-fit} $& \multicolumn{2}{c|}{3017.12}& \multicolumn{2}{|c|}{3018.32}\\
			\hline
		\end{tabular}
		\label{tab:2}
	\end{center}
\end{table}

\begin{table}[h]
	\begin{center}
		\caption{Fitting results for the TECE and $\Lambda$CDM models, where the limits are given at 68$\%$ and 95$\%$ C.L., respectively }
		\begin{tabular} {|c|c|c|c|c|}
			\hline
			Parameter & TECE (68\% C.L.)& TECE (95\% C.L.) & $\Lambda$CDM (68\% C.L.)& $\Lambda$CDM (95\% C.L.)\\
			\hline
		
			{\boldmath$\Omega_b h^2   $} & $0.02226\pm 0.00016        $& $0.02226^{+0.00033}_{-0.00032}$ & $0.02236\pm 0.00014        $& $0.02236^{+0.00028}_{-0.00027}$\\
			
			{\boldmath$\Omega_c h^2   $} & $0.1174\pm 0.0013          $& $0.1174 \pm 0.0025$ & $0.1173\pm 0.0012          $& $0.1173 \pm 0.0023$\\
			
			{\boldmath$100\theta_{MC} $} & $1.04125\pm 0.00036        $& $1.04125^{+0.00073}_{-0.00069}$& $1.04099\pm 0.00031        $& $1.04099^{+0.00062}_{-0.00060}$\\
			
			{\boldmath$\tau           $} & $0.090\pm 0.021            $& $0.090 \pm 0.042   $  & $0.079\pm 0.018            $ & $0.079^{+0.037}_{-0.035}   $\\
			
			{\boldmath$\Sigma m_\nu   $/eV} & $< 0.186                   $& $< 0.317                   $  & $< 0.107                   $& $< 0.195                   $\\
			
			{\boldmath$\xi            $} & $0.00038\pm 0.00027        $& $0.00038^{+0.00055}_{-0.00053}$&$-$&$-$\\
			
			{\boldmath${\rm{ln}}(10^{10} A_s)$} & $3.103\pm 0.039            $& $3.103 \pm 0.076   $& $3.085\pm 0.034            $& $3.085^{+0.068}_{-0.065}   $\\
			
			$H_0                       $ & $68.42\pm 0.71             $& $68.4 \pm 1.4        $ & $68.05^{+0.60}_{-0.54}     $ & $68.1^{+1.1}_{-1.2}        $\\
			
			$\sigma_8                  $ & $0.814^{+0.017}_{-0.014}   $& $0.814^{+0.028}_{-0.032}   $& $0.814^{+0.014}_{-0.012}   $ & $0.814^{+0.024}_{-0.027}   $\\
			\hline
			$\chi^2_{best-fit} $& \multicolumn{2}{c|}{3018.96}& \multicolumn{2}{|c|}{3019.28}\\
			\hline
		\end{tabular}
		\label{tab:3}
	\end{center}
\end{table}

In Fig.~\ref{fg:P}, we present our fitting results of  PLCE (red) and $\Lambda$CDM (blue).
Although the PLCE model has been discussed in the literature, it is the first time to illustrate its numerical cosmological effects.
In particular, we find that $\nu$ = $( 0.0240^{+0.0110}_{-0.0085})$ in 68$\%$~C.L., which shows that PLCE and $\Lambda$CDM can be clearly distinguished.
It is interesting to note that the value of $\sigma_8$=$0.814^{+0.023}_{-0.026} $ (95$\%$~C.L.) in PLCE
is smaller than that of $0.815^{+0.023}_{-0.025}$ (95$\%$~C.L.) in $\Lambda$CDM. 
As shown in  Table~\ref{tab:2}, the best fitted $\chi^2$ value in PLCE is 3017.12, which is also smaller than 3018.32 in $\Lambda$CDM.
Although the cosmological observables for the best $\chi^2$ fit in PLCE do not significantly deviate from those in $\Lambda$CDM,
it indicates that the PLCE model is closer to the observational data than $\Lambda$CDM.

Similarly, we show our results for TECE (red) and $\Lambda$CDM (blue)
 in Fig.~\ref{fg:T}. Explicitly, we get that $\xi$ = $( 3.8\pm {2.7})\times 10^{-4}$ 
in 68$\%$~C.L.
In addition, the TECE model can relax the limit of the total mass of the active neutrinos.
In particular, we have that $\Sigma m_\nu$ $< 0.317$ eV, comparing to $\Sigma m_\nu$ $< 0.195$ eV in $\Lambda$CDM at 95$\%$~C.L.
In addition, the value of $H_0$ in TECE equals to $68.42\pm {0.71}$ $(68.4\pm {1.4})$, which is larger than $68.05^{+0.60}_{-0.54}$ $(68.1^{+1.1}_{-1.2})$ in $\Lambda$CDM with 68$\%$ (95$\%$)~C.L.

As shown in Table \ref{tab:3}, the best fitted $\chi^2$ value in the TECE
model is 3018.96, which is smaller than 3019.28 in $\Lambda$CDM model.
Although the difference between the value of $\chi^2$ in TECE and $\Lambda$CDM is not significant,
it still implies that the TECE model can not be ignored. Clearly, more considerations and discussions are needed in the future.

\section{Conclusions}

We have calculated the cosmological evolutions of $\rho_{DE}$ and $w_{DE}$ in the PLCE and TECE models.
We have found that the EoS of dark energy in PLCE (TECE) does not cross -1.
We have shown that the CMB TE power spectrum of the PLCE model with a positive $\nu$ is closer to the Planck 2018 data than that in $\Lambda$CDM, while the CMB TT spectrum in the TECE model has  smaller values around $l\sim 20-27$, which are lower than that in $\Lambda$CDM, but  close to the data of Planck 2018.
By using the Newton method in the global fitting, we have obtained the first numerical result in the PLCE model with $\nu=0.0240^{+0.0110}_{-0.0085} $
in 68$\%$ C.L., which can be distinguished well with $\Lambda$CDM. Our Fitting results indicate that the PLCE model gives a smaller value of $\sigma_8$ with a better $\chi^2$ value than $\Lambda$CDM.
In the TECE model, we have gotten that $\xi=(3.8\pm2.7)\time 10^{-4}$ and $\Sigma m_\nu$ $< 0.186$ eV in 68$\%$ C.L., while $H_0$ is closer to 70. 
The best fitted value of $\chi^2$ is 3018.96 in the TECE model, which is smaller than 3019.28 in $\Lambda$CDM. These results have demonstrated that the TECE model deserves more attention and research in the future.

\begin{acknowledgments}
This work was supported in part by National Center for Theoretical Sciences and
MoST (MoST-107-2119-M-007-013-MY3).
\end{acknowledgments}



\begin{thebibliography}{99}
\bibitem{Amendola:2015}
L. Amendola and S. Tsujikawa, \textit{Dark Energy : Theory and Observations}, (Cambridge Univer-
sity Press, 2015).	

\bibitem{Weinberg:1988cp} 
S.~Weinberg,
Rev.\ Mod.\ Phys.\  {\bf 61}, 1 (1989).

\bibitem{Weinberg:1972} 
S.~Weinberg,
\textit{Gravitation and Cosmology}, (Wiley and Sons, New York, 1972).

\bibitem{ArkaniHamed:2000tc} 
N.~Arkani-Hamed, L.~J.~Hall, C.~F.~Kolda and H.~Murayama,
Phys.\ Rev.\ Lett.\  {\bf 85}, 4434 (2000).

\bibitem{Peebles:2002gy} 
P.~J.~E.~Peebles and B.~Ratra,
Rev.\ Mod.\ Phys.\  {\bf 75}, 559 (2003).

\bibitem{Copeland:2006wr} 
E.~J.~Copeland, M.~Sami and S.~Tsujikawa,
Int.\ J.\ Mod.\ Phys.\ D {\bf 15}, 1753 (2006).

\bibitem{Jacobson:1995ab} 
T.~Jacobson,
Phys.\ Rev.\ Lett.\  {\bf 75}, 1260 (1995).

\bibitem{Cai:2005ra} 
R.~G.~Cai and S.~P.~Kim,
JHEP {\bf 0502}, 050 (2005).


\bibitem{Akbar:2006er} 
M.~Akbar and R.~G.~Cai,
Phys.\ Lett.\ B {\bf 635}, 7 (2006)

\bibitem{Akbar:2006kj} 
M.~Akbar and R.~G.~Cai,
Phys.\ Rev.\ D {\bf 75}, 084003 (2007)

\bibitem{Jamil:2009eb} 
M.~Jamil, E.~N.~Saridakis and M.~R.~Setare,
Phys.\ Rev.\ D {\bf 81}, 023007 (2010)

\bibitem{Fan:2014ala} 
Z.~Y.~Fan and H.~Lu,
Phys.\ Rev.\ D {\bf 91}, no. 6, 064009 (2015)

\bibitem{Gim:2014nba} 
Y.~Gim, W.~Kim and S.~H.~Yi,
JHEP {\bf 1407}, 002 (2014)

\bibitem{Das:2007mj}
S.~Das, S.~Shankaranarayanan and S.~Sur,
Phys.\ Rev.\ D {\bf 77}, 064013 (2008).

\bibitem{Tsallis:1987eu} 
C.~Tsallis,
J.\ Statist.\ Phys.\  {\bf 52}, 479 (1988).



\bibitem{Lyra:1998wz} 
M.~L.~Lyra and C.~Tsallis,
Phys.\ Rev.\ Lett.\  {\bf 80}, 53 (1998).

\bibitem{Wilk:1999dr} 
G.~Wilk and Z.~Wlodarczyk,
Phys.\ Rev.\ Lett.\  {\bf 84}, 2770 (2000).

\bibitem{Easson:2010av} 
D.~A.~Easson, P.~H.~Frampton and G.~F.~Smoot,
Phys.\ Lett.\ B {\bf 696}, 273 (2011).


\bibitem{Komatsu:2013qia} 
N.~Komatsu and S.~Kimura,
Phys.\ Rev.\ D {\bf 88}, 083534 (2013).

\bibitem{Abreu:2014ara} 
E.~M.~C.~Abreu {\it et al.}, 
Physica A {\bf 441}, 141 (2016);
S.~Ghaffari {\it et al.}, 
Eur.\ Phys.\ J.\ C {\bf 78},  706 (2018);
 M.~A.~Zadeh, A.~Sheykhi, H.~Moradpour and K.~Bamba,
Eur.\ Phys.\ J.\ C {\bf 78},  940 (2018);
E.~N.~Saridakis, K.~Bamba, R.~Myrzakulov and F.~K.~Anagnostopoulos,
JCAP {\bf 1812}, 012 (2018);
 M.~Tavayef, A.~Sheykhi, K.~Bamba and H.~Moradpour,
Phys.\ Lett.\ B {\bf 781}, 195 (2018).


\bibitem{Nunes:2014jra} 
E.~M.~Barboza {\it et al.},
Physica A {\bf 436}, 301 (2015).


\bibitem{Zadeh:2018wub} 
M.~Abdollahi Zadeh, A.~Sheykhi and H.~Moradpour,
Mod.\ Phys.\ Lett.\ A {\bf 34},  1950086 (2019).

\bibitem{Nojiri:2019skr} 
S.~Nojiri, S.~D.~Odintsov and E.~N.~Saridakis,
Eur.\ Phys.\ J.\ C {\bf 79}, no. 3, 242 (2019).

\bibitem{Tsallis:2012js} 
C.~Tsallis and L.~J.~L.~Cirto,
Eur.\ Phys.\ J.\ C {\bf 73}, 2487 (2013).

\bibitem{1205.3421}
K.~Bamba, S.~Capozziello, S.~Nojiri and S.~D.~Odintsov,
  Astrophys.\ Space Sci.\  {\bf 342}, 155 (2012).


\bibitem{Lymperis:2018iuz} 
A.~Lymperis and E.~N.~Saridakis,
Eur.\ Phys.\ J.\ C {\bf 78}, 993 (2018).


\bibitem{press:2007}
W.~H.~Press {\it et al.}, 
{\it Numerical Recipes 3rd Edition: The Art of
	Scientific Computing}, (Cambridge University Press, 2007). 

\bibitem{Lewis:2002ah}
A.~Lewis and S.~Bridle,
Phys.\ Rev.\ D {\bf 66}, 103511 (2002).



\bibitem{Zhang:2012mp} 
C.~Zhang, H.~Zhang, S.~Yuan, T.~J.~Zhang and Y.~C.~Sun,
Res.\ Astron.\ Astrophys.\  {\bf 14},  1221 (2014).

\bibitem{Jimenez:2003iv} 
R.~Jimenez, L.~Verde, T.~Treu and D.~Stern,
Astrophys.\ J.\  {\bf 593}, 622 (2003).

\bibitem{Simon:2004tf} 
J.~Simon, L.~Verde and R.~Jimenez,
Phys.\ Rev.\ D {\bf 71}, 123001 (2005).

\bibitem{Moresco:2012jh} 
M.~Moresco {\it et al.},
JCAP {\bf 1208}, 006 (2012).

\bibitem{Gaztanaga:2008xz} 
E.~Gaztanaga, A.~Cabre and L.~Hui,
Mon.\ Not.\ Roy.\ Astron.\ Soc.\  {\bf 399}, 1663 (2009).

\bibitem{Moresco:2016mzx} 
M.~Moresco {\it et al.},
JCAP {\bf 1605},  014 (2016).

\bibitem{Stern:2009ep} 
D.~Stern, R.~Jimenez, L.~Verde, M.~Kamionkowski and S.~A.~Stanford,
JCAP {\bf 1002}, 008 (2010).

\bibitem{Reid:2012sw} 
B.~A.~Reid {\it et al.},
Mon.\ Not.\ Roy.\ Astron.\ Soc.\  {\bf 426}, 2719 (2012).

\bibitem{Moresco:2015cya} 
M.~Moresco,
Mon.\ Not.\ Roy.\ Astron.\ Soc.\  {\bf 450},  L16 (2015).

\bibitem{Busca:2012bu} 
N.~G.~Busca {\it et al.},
Astron.\ Astrophys.\  {\bf 552}, A96 (2013).

\bibitem{Hu:2014vua} 
Y.~Hu, M.~Li and Z.~Zhang,
arXiv:1406.7695 [astro-ph.CO].

\bibitem{Font-Ribera:2013wce} 
A.~Font-Ribera {\it et al.} [BOSS Collaboration],
JCAP {\bf 1405}, 027 (2014).






\bibitem{Adam:2015wua} 
R.~Adam {\it et al.} [Planck Collaboration],
Astron.\ Astrophys.\  {\bf 594}, A10 (2016).


\bibitem{Aghanim:2015xee} 
N.~Aghanim {\it et al.} [Planck Collaboration],
Astron.\ Astrophys.\  {\bf 594}, A11 (2016).

\bibitem{Ade:2015zua} 
P.~A.~R.~Ade {\it et al.} [Planck Collaboration],
Astron.\ Astrophys.\  {\bf 594}, A15 (2016).

\bibitem{Heymans:2013fya}
  C.~Heymans {\it et al.},
Mon.\ Not.\ Roy.\ Astron.\ Soc.\  {\bf 432}, 2433 (2013).

\bibitem{Beutler:2011hx}
F.~Beutler {\it et al.},
Mon.\ Not.\ Roy.\ Astron.\ Soc.\  {\bf 416}, 3017 (2011).

\bibitem{Anderson:2013zyy}
L.~Anderson {\it et al.} [BOSS Collaboration],
Mon.\ Not.\ Roy.\ Astron.\ Soc.\  {\bf 441}, 24 (2014).


\end{thebibliography}
\end{document}